\documentclass[11pt]{article}
\usepackage{epsfig}
\usepackage[T1]{fontenc}
\newcommand{\be}{\begin{equation}}
\newcommand{\en}{\end{equation}}
\newcommand{\bea}{\begin{eqnarray}}
\newcommand{\ena}{\end{eqnarray}}
\newcommand{\lbl}[1]{\label{eq:#1}}
\newcommand{\lbltab}[1]{\label{tab:#1}}
\newcommand{\lblfig}[1]{\label{fig:#1}}
\newcommand{\lblsec}[1]{\label{sec:#1}}
\newcommand{\rf}[1]{(\ref{eq:#1})}
\newcommand{\Table}[1]{\ref{tab:#1}}
\newcommand{\fig}[1]{\ref{fig:#1}}
\newcommand{\sect}[1]{\ref{sec:#1}}
\newcommand{\braque}[1]{{\langle #1 \rangle}}
\newcommand{\gapprox}{%
\mathrel{%
\setbox0=\hbox{$>$}\raise0.6ex\copy0\kern-\wd0\lower0.65ex\hbox{$\sim$}}}
\newcommand{\lapprox}{%
\mathrel{%
\setbox0=\hbox{$<$}\raise0.6ex\copy0\kern-\wd0\lower0.65ex\hbox{$\sim$}}}
\newcommand{\inleft}{%
\mathrel{%
\setbox0=\hbox{$<$}\copy0\kern-0.5\wd0\lower1.1\ht0\hbox{$\scriptstyle{in}$}}}
\newcommand{\inright}{%
\mathrel{%
\setbox0=\hbox{$>$}\copy0\kern-0.5\wd0\lower1.1\ht0\hbox{$\scriptstyle{in}$}}}
\newcommand{\outleft}{%
\mathrel{%
\setbox0=\hbox{$<$}\copy0\kern-0.5\wd0\lower1.1\ht0\hbox{$\scriptstyle{out}$}}}
\newcommand{\outright}{%
\mathrel{%
\setbox0=\hbox{$>$}\copy0\kern-0.5\wd0\lower1.1\ht0\hbox{$\scriptstyle{out}$}}}

\newcommand{\im}{{\rm Im}}
\newcommand{\smatrix}{{$S$-matrix\ }}

\newcommand{\mkd} {m^2_K}

\newcommand{\mpid} {m^2_\pi}

\newcommand{\mtaud} {m^2_{\tau}}

\def\im{ {\rm Im}\,}

\def\tr{ {\rm tr}\,}
\def\Lag{ {\cal L}}

\def\D{ {\cal D}}
\begin{document}

\begin{center}
{\Large\bf Analyticity constraints on the strangeness changing vector
current and applications to $\tau\to K\pi \nu_\tau$, 
$\tau\to K\pi\pi \nu_\tau$
}

\bigskip
{\large B. Moussallam
\footnote{\tt moussall@ipno.in2p3.fr}
}

{\sl Groupe de Physique Th\'eorique, 
Institut de Physique Nucl\'eaire, CNRS/IN2P3,\\ 
Universit\'e Paris-Sud 11,
91406 Orsay, France}
\bigskip

{\Large\bf Abstract}\\[5pt]
\begin{minipage}{0.8\textwidth}
We discuss matrix elements of the strangeness changing vector current 
using their relation, due to analyticity, with $\pi K$ scattering in the
$P$-wave.
We take into account 
experimental phase-shift  measurements in the elastic channel
as well as results, obtained by the LASS collaboration, on the details of
inelastic scattering, which show the dominance of two quasi two-body channels
at medium energies. 
The associated form factors are shown to be completely determined, up to
one flavour symmetry breaking parameter, imposing boundary conditions at
$t=0$ from chiral and flavour symmetries and at $t\to\infty$ from QCD.
We apply the results to the $\tau\to K\pi \nu_\tau$ 
and $\tau\to K\pi\pi \nu_\tau$ amplitudes 
and compare the former to recent high statistics results from B factories.
\medskip

\end{minipage}
\end{center}
\bigskip

\section{Introduction}
Decays of the $\tau$  
to hadrons with strangeness $S=-1$ can be used to determine basic parameters
of the standard model such as $V_{us}$ 
and the mass of the
strange quark (e.g.~\cite{pichvus,maltman} for recent updates). 
The considerable improvement in statistics
brought in by Babar and Belle should translate in the near future into much
more precise measurements of matrix elements of currents with $S=-1$ than
possible at present. In this paper, we re-consider in some detail 
the relations between the $K\pi$ matrix element of the strangeness-changing
vector current and $K\pi$ scattering in the $P$-wave. 

We will follow a method applied some time ago by Donoghue, Gasser and 
Leutwyler (DGL~\cite{dgl})  to the $\pi\pi$ matrix element of the
$S=0$ scalar current (which is not directly accessible to experiment
since the Higgs boson is not very light).
For this method to operate, it is necessary that
inelastic scattering can be approximated in terms of a finite number 
of two-body or quasi-two-body channels in a sufficiently large energy range
$E\lapprox E_0$. Then, imposing a limited number of
constraints at $E=0$ from chiral symmetry
and at $E >> E_0$ from asymptotic QCD one can determine the form factor
from the $T$-matrix. In practice, this is done by solving a set of coupled
Muskhelishvili-Omn\'es (MO) integral equations which are consequences of
analyticity properties of the form factors and of time reversal invariance.

Recently, this method was implemented in the case of the 
$K\pi$ scalar form factor~\cite{jop}: in this case, inelasticity
in the $S$-wave is saturated by the $K\eta$ and $K\eta'$ channels. 
One motivation for our interest
in the vector form factor is the availability of good experimental data
from the LASS collaboration
on both elastic~\cite{aston88} and inelastic $K\pi$ 
scattering~\cite{aston84,aston87,aston88b} in the $P$-wave. 
In particular, these works show that in an energy  range 
$E\lapprox 2.5$ GeV, inelasticity is dominated by two
quasi-two-body channels: $K^*\pi$ and $K\rho$. 
This makes it possible to probe the DGL method by comparing its results
with experimental data from $\tau$ decays. The spectral function for
$\tau\to K\pi \nu_\tau$ can be described, of course, by a simple
Breit-Wigner ansatz in the vicinity of the $K^*(892)$ resonance, where
scattering is nearly perfectly elastic. Away from the resonance, however,
this is no longer true. In the energy region $E<< M_{K^*}$  the spectral
function gets dominated by the scalar component~\cite{joptau} 
(as it is less suppressed by phase-space than the vector component). In the
energy region $E>> M_{K^*}$ it is important to correctly treat inelastic
effects. 
Another motivation for this work is in view of applications to three body
non-leptonic $B$-decays $B\to K\pi\pi$. In kinematical configurations where
the kaon and one pion are quasi aligned, factorization can presumably 
be justified 
\cite{benekeparis} and the amplitude gets expressed in terms of $K\pi$
vector and scalar form factors\cite{elbennich}\footnote{
In the context of B decays, the usefulness of appealing to 
descriptions more sophisticated
than Breit-Wigner combinations for scalar form factors
was pointed out in ref.~\cite{gardnermeissner}.}.

The plan of the paper is as follows. After introducing some notation
for the form factors involved we discuss the construction of the
$T$-matrix from fits to the experimental $\pi K$ scattering data. 
Three-channel unitarity is enforced using the $K$-matrix method, and proper
flavour symmetry and low energy behaviour are enforced starting from
a resonance chiral Lagrangian. Next, we introduce the MO integral equations
satisfied by the form factors and discuss their resolution. This
necessitates to make a plausible ansatz for the $T$-matrix at high
energies. The ansatz determines the number of conditions to be imposed
in order to determine the form factors from solving the integral
equations. We will use three conditions at $E=0$ and one asymptotic 
condition. Finally, the results are displayed and compared with
recent experimental measurements.
\section{Strangeness changing vector current form factors}
The object which will mainly interest us is the $K\pi$ matrix element
of the strangeness changing vector current
\bea\lbl{def1}
&& <K^+(p_K)\vert \bar{u}\gamma^\mu s\vert \pi^0(p_\pi)>=
\nonumber\\
&& \quad f_+^{K^+\pi^0}(t)\,(p_K+p_\pi)^\mu +
f_-^{K^+\pi^0}(t)\,(p_K-p_\pi)^\mu \ ,
\ena
with $t=(p_K-p_\pi)^2$ and we will denote the vector form factor as
\be
H_1(t)\equiv f_+(t)=\sqrt2 f_+^{K^+\pi^0}(t)\ .
\en
As we will discuss below $K\pi$ inelastic scattering in the $P$-wave 
is dominated by two quasi two-body channels: $K^*\pi$ and $\rho K$. This
leads us to introduce the associated vector current matrix elements 
\bea
&&\braque{K^{*+}(p_V,\lambda) \vert \bar{u}\gamma_\mu s\vert \pi^0(p_\pi)}=
\epsilon_{\mu\nu\alpha\beta}\, e^{*\nu}(\lambda) p_V^{\alpha} p_{\pi}^{\beta}\,
H_2(t) \qquad \nonumber \\
&&\braque{\rho^0(p_V,\lambda) \vert \bar{u}\gamma_\mu s\vert K^-(p_K)}=
-\epsilon_{\mu\nu\alpha\beta}\, e^{*\nu}(\lambda) p_V^{\alpha} p_{K}^{\beta}\,
H_3(t)  \nonumber \\
\ena
(we have chosen different signs in the definition of $H_2$ and $H_3$
such that only plus signs appear in subsequent equations). 
In the following,
isospin symmetry breaking will be neglected. From the isospin decomposition
of $K^+\pi^0$ it  follows that
\be
\braque{K^+\pi^0\vert \bar{u}\gamma_\mu s\vert0}=
\sqrt{{1\over3}} \braque{[K\pi]_{{1\over2}}  \vert \bar{u}\gamma_\mu s\vert0}
\en
and analogous relations hold for $K^*\pi$, $K\rho$. In order to derive
the unitarity equations satisfied by the form factors $H_1$, $H_2$ and $H_3$
it is convenient to focus on one of the spatial components of the
vector current, say $\mu=3$, and go to center-of-mass (CMS) frame of the 
meson pair. In this frame, 
the current matrix elements introduced above can be expressed as follows
\be\lbl{j3kpi}
\braque{[K\pi]_{{1\over2}}\vert \bar{u}\gamma^3 s\vert0}=
\sqrt6\,\cos\theta_0\,q_{K\pi}(t)\,H_1(t)
\en
where $q_{K\pi}$ is the modulus of the CMS momentum of the meson pair
and $\theta_0$ is the polar angle of the momentum vector with respect
to a fixed frame. Similarly, the matrix elements which involve one
vector meson read
\bea\lbl{j3vp}  
&& \braque{ [K^*(\lambda)\pi]_{{1\over2}}\vert \bar{u}\gamma^3 s  \vert0}
=i\eta^* \sqrt{3\over2}\, \sin\theta_0 \sqrt{t}\,  q_{ K^*\pi}(t)\,  H_2(t)
\qquad \nonumber \\
&&\braque{ [\rho (\lambda) K]_{{1\over2}}\vert \bar{u}\gamma^3 s  \vert0}
\phantom{*}=
i\eta^* \sqrt{3\over2}\, \sin\theta_0 \sqrt{t}\,  q_{ \rho K}(t)\,  H_3(t)\ .
\ena
Here, $\eta$ is an arbitrary phase which can be introduced in the
definition of the vector meson polarization vector 
\bea\lbl{polarphas} 
&& {\vec{e^*}}(\lambda=\pm1) 
={\eta^*\over\sqrt2} \left(
\begin{array}{c}
-\lambda\cos\theta_0\cos\phi_0 -i\sin\phi_0 \\
-\lambda\cos\theta_0\sin\phi_0 +i\cos\phi_0 \\
 \lambda\sin\theta_0
\end{array}
\right)
\ena
($\theta_0,\ \phi_0$ being the polar angles of the vector meson momentum). 
In the following, we will set $i\eta^*=1$. We have also taken the following 
convention for the Levi-Civitta tensor
\be
\epsilon_{0123}=1\ .
\en
The explicit dependence on $\theta_0$ displayed above in 
eqs.~\rf{j3kpi}~\rf{j3vp} indicates that these matrix elements concern 
the angular momentum state $J=1$ of the meson pair.

\section{$K\pi$ scattering in $P$-wave with $n$-channel unitarity}

\subsection{Experimental situation}
Detailed partial-wave analysis of $K\pi\to K\pi$ scattering  (in the
energy range $E\lapprox 2.5$ GeV) have been performed based on high
statistics production experiments $K p \to K \pi N$ in 
refs.~\cite{estabrooks,aston88}. The LASS collaboration has also performed
production experiments $K^-p \to K\,2\pi N$ 
and $K^-p \to K\,3\pi N$~\cite{aston84,aston87,aston88b} which provide 
informations on inelastic $K\pi$ scattering. These, however, are not as 
precise as for elastic scattering and mainly concern resonance properties
in the various partial-waves. 
%
%
In the $S$-wave, $K\pi$ scattering  is elastic to a good approximation 
up to the $K\eta'$ threshold\cite{estabrooks,aston88}, 
indicating that $K\eta'$ is the main inelastic
channel. Inelasticity in the $P$-wave starts at somewhat smaller energy
than in the $S$-wave and the results of 
refs.~\cite{aston84,aston87} suggest that $K^*\pi$ 
is the main inelastic channel, as it is enhanced by two resonances
$K^*(1410)$ and $K^*(1680)$ (following the nomenclature of 
the PDG~\cite{pdg06} ). The latter resonance was observed to couple 
essentially to three channels: $K\pi$, $K^*\pi$ as well as 
$K\rho$~\cite{pdg06}. 
This indicates that $K\rho$ should be also taken into account as a significant
inelastic channel in the $P$-wave although its coupling to $K^*(1410)$
was found to be small. The experimental results on the branching ratios
of the $K^*(1410)$ and $K^*(1680)$ are collected in table ~\Table{resodecay}.

The amplitude $K\pi\to K\eta$ was studied 
in ref.\cite{aston88b} and found to be very small in the $P$-wave 
and not to display any resonant effect. The coupling of $K\eta'$ to 
resonances with $J^{PC}=1^{--}$ is proportional to the sine of the 
mixing angle and thus should also be suppressed. 
These results suggest that a 
plausible  description of $K\pi$ scattering in the $P$-wave with
$n$-channel unitarity can be performed (up to energies $E\lapprox 2.5$ GeV)
by retaining as inelastic channels the two 
quasi two-body channels with one vector meson: $K^*\pi$ and $K\rho$.
\begin{table}[hbt]
\centering
\begin{tabular}{|l|lll|}\hline
\           & $K\pi $      & $K^*\pi$             & $ K \rho$ \\ \hline
$K^*(1410)$ & $6.6\pm 1   $& $>40   $             & $ < 7  $  \\
$K^*(1680)$ & $38.7\pm2.5 $& $29.9^{+2.2}_{-4.7}$ & $31.4^{+4.7}_{-2.1}$ 
\\ \hline
\end{tabular}
\caption{\sl Experimental decay branching ratios (in percent units)
for the resonances $K^*(1410)$ and $K^*(1680)$ quoted in the PDG.}
\lbltab{resodecay}
\end{table}
\subsection{Lagrangian for vector resonances and pseudo-scalar mesons}
It is useful to express the resonance contribution to scattering in terms
of coupling constants which are known in other contexts and it is also
useful to impose chiral constraints at low energy. 
For this reason, let us start from the following Lagrangian which was used
in the context of the chiral expansion\cite{eglpr,prades93}. It includes
the nonet of the light vector mesons encoded in a matrix $V_\mu$ and the
light pseudoscalars encoded in a chiral matrix $u_\mu$ and involves 
two coupling constants $g_V$ and $\sigma_V$,
\be
\Lag^{(1)}=\Lag^{(1)}_K + \Lag^{(1)}_V+\Lag^{(1)}_{\sigma}
\en  
with
\bea\lbl{lagv1}
&& \Lag^{(1)}_K = {-1\over4}\tr( V_{\mu\nu}  V^{\mu\nu} -2 M_V^2 V_\mu V^\mu),
\nonumber\\
&& \Lag^{(1)}_V = {-i\over2} g_V\, \tr( V_{\mu\nu} [u_\mu,u_\nu])
\nonumber\\
&& \Lag^{(1)}_{\sigma} ={1\over 2} \sigma_V\, \epsilon^{\mu\nu\rho\sigma} 
\tr( V_\mu \{ u_\nu, V_{\rho\sigma}\} )
\ena
and $V_{\mu\nu}=\nabla_\mu V_\nu -\nabla_\nu V_\mu$.
In ref.\cite{prades93}, for example, the following values are quoted 
for the coupling constants (based on the extended NJL model)
\be
g_V \simeq 0.083,\quad \sigma_V\simeq 0.25  
\en
which should serve as a guide as to the orders of magnitudes. Let us denote
an excited vector resonance by $V_\mu ^{(n)}$, we can write down the
following coupling terms
\bea\lbl{lagvn}
&& \Lag^{(n)}_V = {-i\over2} g_V(n)\, \tr( V_{\mu\nu}^{(n)} [u_\mu,u_\nu])
\nonumber\\
&& \Lag^{(n)}_{\sigma} ={1\over 2} \sigma_V(n)\, \epsilon^{\mu\nu\rho\sigma} 
\tr( V_\mu^{(n)} \{ u_\nu, V_{\rho\sigma}\} )\ .
\ena
These terms do not involve the quark mass matrix and, therefore, 
have exact $SU(3)$ flavour symmetry. Experimental results on $K^*$
resonances indicate that flavour symmetry can sometimes be 
substantially broken, as in  the case of  the $K^*(1410)$ 
(see table \Table{resodecay}). 
We do not try to write down all possible Lagrangian
terms which break flavour symmetry  but eventually will implement 
such effects at the level of the fits. We also do not consider explicitly the
possibility of many more terms which involve further derivatives acting
on the vector or on the chiral fields. Again, such terms which give
rise to polynomial energy dependence, may be implemented phenomenologically
as ``background'' contributions in the fits.  
\subsection{Resonance contributions to $K\pi$ scattering}
Let us recall the constraints introduced by the conservation of parity. The
$K\pi $ system, at first, in the $P$-wave has parity $-1$.
The action of the parity operator on a 
state formed from a vector meson and a pseudo-scalar
meson involves  the helicity\cite{jacobwick},
\be
{\cal P} \vert V(\lambda) P  >_J = (- )^{J-1} \vert V(-\lambda) P>_J\ .
\en
For $J=1$, the following combination is the only one which has negative parity 
\be\lbl{pmincombo}
\psi_- = {1\over\sqrt2}\left( \vert V(1) P  >_{J=1}
- \vert V(-1) P  >_{J=1}\right)\ .
\en
We will need the partial-wave expansion of the scattering amplitudes
in a situation where the initial state CMS momentum has polar angles
$\theta_0,\ \phi_0$ and the final state has polar 
angles $\theta,\ \phi$~\cite{jacobwick}
\bea\lbl{Tpwexp}
&&   \braque{ V(\lambda) P\vert T\vert K\pi} =
16\pi \sum_{J,M} (2J+1)\,(q_{VP}\, q_{K\pi} )^J\times 
\nonumber\\ 
&&\quad  \braque{ V(\lambda) P\vert T^J\vert K\pi}
\D^{*J}_{M,\lambda}(\phi,\theta) \D^{J}_{M,0}(\phi_0,\theta_0)
\nonumber\\
&&   \braque{ V(\lambda) P\vert T \vert V'(\lambda') P }=
16\pi \sum_{J,M} (2J+1) \,  (q_{VP}\, q_{V'P} )^J \times
\nonumber\\
&&\quad \braque{V(\lambda) P\vert T^J\vert V'(\lambda') P }
\D^{*J}_{M,\lambda}(\phi,\theta) \D^{J}_{M,\lambda'}(\phi_0,\theta_0)\ .
\nonumber\\
\ena
We have factored out explicitly the angular momentum barrier factors
in the definition of the partial-wave $T$-matrix elements: this is
necessary for non-diagonal matrix elements in order to ensure good
analytical behaviour for the partial-wave $T$-matrix~\cite{frazerfulco}.
For simplicity, we also define the diagonal elements in the same manner
e.g.
\bea
&&\braque{K\pi\vert T\vert K\pi}=
16\pi \sum_{J,M} (2J+1)\,(q_{K\pi})^{2J}  \times
\nonumber\\ 
&& \qquad \braque{K\pi\vert T^J\vert K\pi} 
\D^{*J}_{M,0}(\phi,\theta) \D^{J}_{M,0}(\phi_0,\theta_0)\ .
\ena
Let us number the three relevant channels in our construction as
\be
1 \longrightarrow K\pi,\quad
2 \longrightarrow K^*\pi,\quad
3 \longrightarrow K\rho,
\en
Using the Lagrangian introduced above~\rf{lagv1},~\rf{lagvn} 
it is not difficult to compute
the resonance contributions to the various $T$-matrix elements.
The resulting amplitude can be written in a compact form 
which displays the usual resonance   structure ,
\be\lbl{compactres} 
T^1_{ij,r} =\sum_n\, {g_r(n,i) g_r(n,j) \over M_n^2 - s }
\en
with
\bea\lbl{compactres1}
&& g_r(n,1)= {g_V(n)\over\sqrt{16\pi} }\left( {\sqrt{s}\over F_\pi}\right)^2
\nonumber\\
&& g_r(n,2)= {\sigma_V(n)\over\sqrt{16\pi} } {\sqrt{2s}\over F_\pi}
(1+\delta_{n1})
\nonumber\\
&& g_r(n,3)= {-\sigma_V(n)\over\sqrt{16\pi} } {\sqrt{2s}\over F_\pi}\ .
\ena
The relations between channel 2 and channel 3 matrix elements
are consequences of exact flavour symmetry as implemented in the 
Lagrangians~\rf{lagv1},~\rf{lagvn}.
The energy dependence of these effective coupling constants is reliable only
for small values of $s$ ( $s<< M_{K^*}^2$ ) where chiral symmetry constraints
(as encoded in the construction of the Lagrangian) 
are relevant. In practice, we will implement
a simple cutoff function by replacing
\be\lbl{bncutoff}
\sqrt{s}\longrightarrow {\sqrt{s} \left(1 +b_n {\sqrt{s}\over M_n}\right)
\over 1+ b_n {{s}\over M_n^2} }
\en
in eqs.~\rf{compactres1}.
This cutoff function has the correct behaviour at small $s$ and effectively
replaces $\sqrt{s}$ by the resonance mass $M_n$ otherwise. 

\subsection{$K$-matrix fits to the experimental data}
We may implement $n$-channel unitarity in a simple way, using the $K$-matrix 
method (e.g.~\cite{dalitztuan}). 
Starting from a real, symmetric $K$-matrix we define the $J=1$
$T$-matrix in the following way (suppressing the $J$ superscript),
\be
T = (1+i K \Sigma Q^2 )^{-1} K
\en  
where  $Q^2$ and $\Sigma$ are diagonal matrices
\be\lbl{qetsigdef}
(Q^2)_{ij}=\delta_{ij} q_i^2,\quad
\Sigma_{ij}= \delta_{ij} {2 q_i\over \sqrt{s} }
\en
$q_i$ being the CMS momentum in channel $i$. Indeed, it not difficult to
verify that the \smatrix, which is defined as follows,
\be\lbl{smatdef}
S = 1 +2i \sqrt{\Sigma}\, Q\, T\, Q\, \sqrt{\Sigma}
\en
is unitary
\be
S S^\dagger =1
\en
and encodes the proper $J=1$ angular momentum barrier factors.

\subsubsection{A first simple fit}
\begin{figure}[ht]
\centering
\includegraphics[width=9cm]{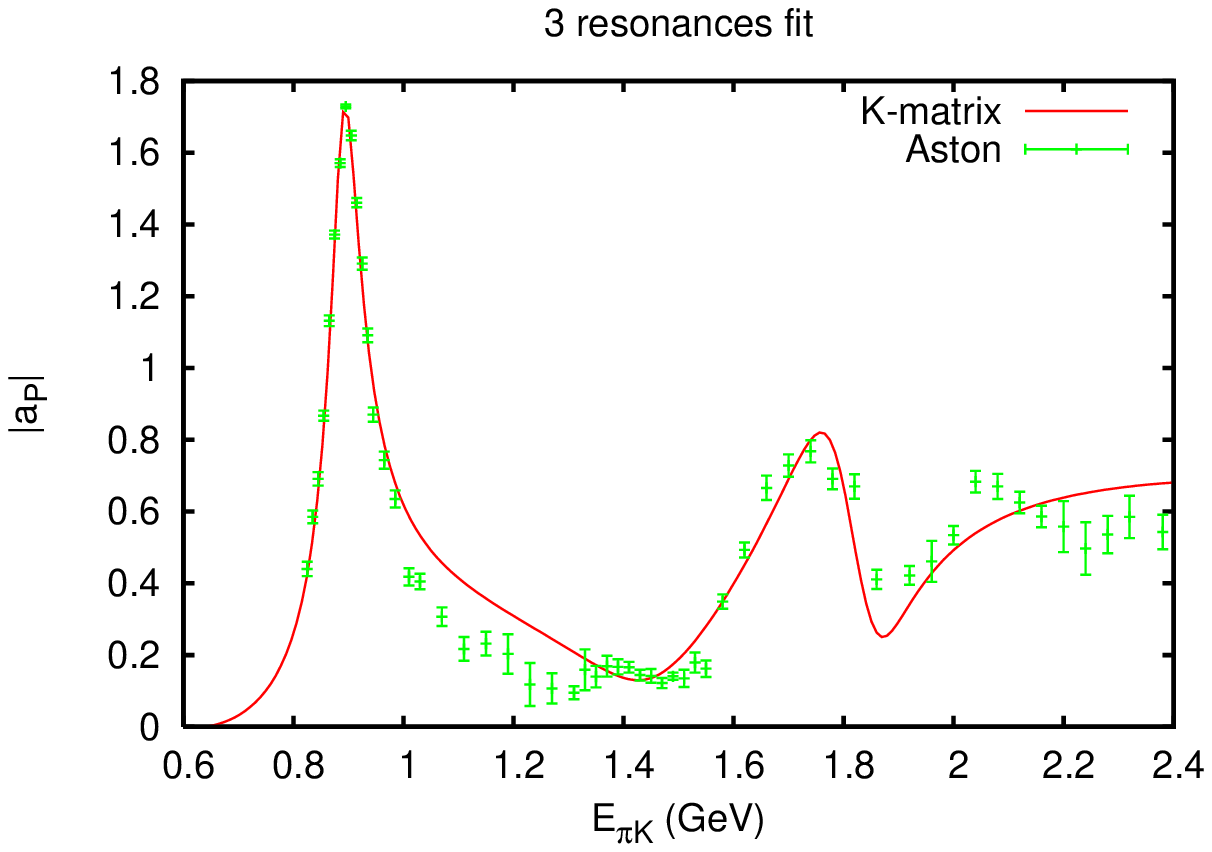}\\
\includegraphics[width=9cm]{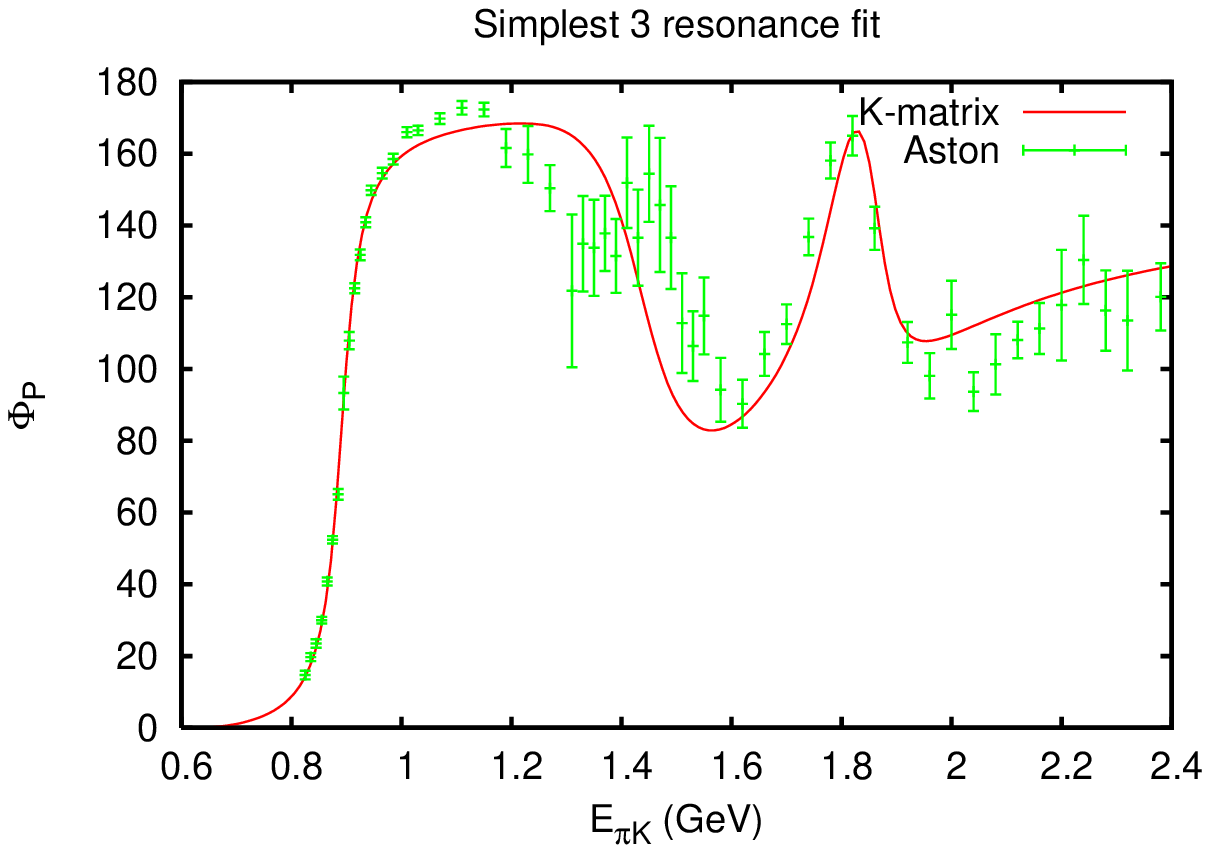}
\caption{\sl $K$-matrix fit to the $K \pi\to K \pi$ data of ref.\cite{aston88}
with three resonances. $a_P$ is the modulus of the $K^-\pi^+\to 
K^-\pi^+$ amplitude  and $\Phi_P$ is the phase.  }
\lblfig{fitsimple}
\end{figure}
The most detailed experimental results concern the $K \pi$ elastic channel.
We used the results of ref.~\cite{aston88} on the amplitudes
$K^- \pi^+ \to K^-\pi^+$ in the energy range $E\le 2.5$ GeV 
and those of ref.~\cite{estabrooks} on the isospin $3/2$ amplitude. 
Let us first perform a simple fit 
including just three resonances in the $K$-matrix, 
i.e., we set $K_{ij}=T^1_{ij,r}$  
using the formulas~\rf{compactres}~\rf{compactres1}. 
We find a qualitatively acceptable result (see figs.~\fig{fitsimple} ) 
with the following resonance parameters,
\be
\begin{array}{ccc}
M_n   & g_V(n)  &\sigma_V(n) \\
0.894\ & 0.0714\  & 0.0989   \\
1.719\ & 0.0103\  &-0.1857   \\
2.247\ & 0.0126\  &-0.3069   \\
\end{array}
\en
(masses are in GeV and the coupling constants are dimensionless). 
In this fit we have taken $b_n=\infty$ (see \rf{bncutoff})
i.e. $\sqrt{s}$ is replaced by $M_n$.
This fit does find a resonance corresponding to $K^*(1680)$ in 
addition to the 
$K^*(892)$  but prefers to locate the third resonance at a higher energy 
rather than at $1.4$ GeV. Obviously, though, the energy region 
between 1 and 1.5 GeV is where the data is not very well described
(the overall $\chi^2/dof =7.4$).

\subsubsection{More sophisticated fits}\lblsec{fitsophi}
Including just one more resonance does not improve the situation very 
much. In order to obtain significantly better fits we will include a fourth
resonance in addition to non-resonant background terms. There are many
physical sources for such terms. We have seen, for instance,  
that the coupling constants may carry energy dependence. In addition, we expect
contributions associated with the left-hand cut, i.e. arising from 
meson exchanges in the crossed channels. We parametrize such contributions
in the following, simplistic, way
\bea
&& K_{11}^{back}= { a_1 s\over 1+ s^3 }
\nonumber\\
&& K_{12}^{back}= { s( a_2 + a_3 s )\over 1+ s^3 }
\nonumber\\
&& K_{13}^{back}= { a_4  s\over 1+ s^3 }\ .
\ena
For the other $K$-matrix elements we do not introduce any background 
dependence. 
We have tried many possibilities but, clearly,
the amount of experimental data is insufficient for probing
in detail all the matrix elements of the $K$ matrix,
so one must make admittedly arbitrary simplifying assumptions. 
We found that setting non-zero background terms for $K_{11}$
$K_{12}$, $K_{13}$ is the most economic way
(in terms of number of parameters) to achieve a good fit. 
In addition to the elastic
$K\pi$ data, we try to reproduce the constraints on the inelastic
channels in the resonance regions (see table \Table{resodecay}). 
These data imply strong
flavour symmetry violation in the region of the $K^*(1410)$ resonance.
We account for this fact by relaxing  
the symmetry relation $g_r(n,3)=-g_r(n,2)$ (see eqs.~\rf{compactres1}) 
for $n=2$. Instead of this relation, 
we suppress the coupling to channel 3 (i.e. $K\rho$ ) by simply setting
\be
g_r(n,3)=0 
\en
when $n=2$. %

Altogether, this fit contains
16 parameters and allows for  a rather satisfactory fit to the 
Aston et al. data, which has a $\chi^2/dof \simeq 1.8$.  
We note that the fourth resonance present in this fit 
effectively acts essentially as an additional
source of background at lower energies: it should not necessarily 
be interpreted as a true physical resonance (in fact no corresponding
resonance is listed in the PDG). 
The numerical results for the best fit parameters are collected in 
table~\Table{fitsophi}. 
\begin{table}[ht]
\centering
\begin{tabular}{|c|c|c|c|c|}\hline
$n$ &   $M_n$     &  $g_V(n)$            & $\sigma_V(n)$   &   $a_n $ \\ \hline
1   &  $0.8962$   & $0.72820\,10^{-1}$   & $0.26080 $       &  $ 3.3906$ \\
2   &  $1.3789$   & $0.52523\,10^{-2}$   & $-0.56075$       &  $ 18.373$ \\
3   &  $1.7300$   & $0.69365\,10^{-2}$   & $-0.21774$       &  $-9.2048$ \\
4   &  $2.2739$   & $0.12044\,10^{-2}$   & $-0.29360$       &  $ 2.8318$ \\ \hline
\end{tabular}
\caption{\sl Results for the fit parameters concerning the four resonances and
the background as described in sec.~\sect{fitsophi}}
\lbltab{fitsophi}
\end{table}

One observes that the values of $g_V(1)$ and $\sigma_V(1)$
are in reasonable agreement with the ENJL predictions 
from ref.~\cite{prades93}. 
Fig.~\fig{fitsophi} shows the comparison of the
fit results with the experimental data. The improvement with respect to the
result of the simpler fit shown in fig.~\fig{fitsimple} is obvious, 
particularly in the energy region [1,1.4] GeV. 
Moreover this fit is able
to reproduce, qualitatively at least, the experimental results concerning
the inelastic channels recalled in table \Table{resodecay}. 
This is illustrated in
fig.~\fig{Sij} which shows the moduli of the transition $S$-matrix elements
and, for the elastic channel $1-S_{11}$ (the cross-sections are proportional
to the squares of these quantities).  Indeed, the figure shows that in the
energy region corresponding to the $K^*(1410)$ the matrix element 
$S_{12}$ shows a clear resonance peak while $S_{13}$ shows no peak. In the
energy region of the $K^*(1680)$ resonance these two matrix elements display
peaks of the same height (this was imposed as a constraint in the fit) and
the peak in the elastic channel is slightly higher.
\begin{figure}[ht]
\centering
\includegraphics[width=9cm]{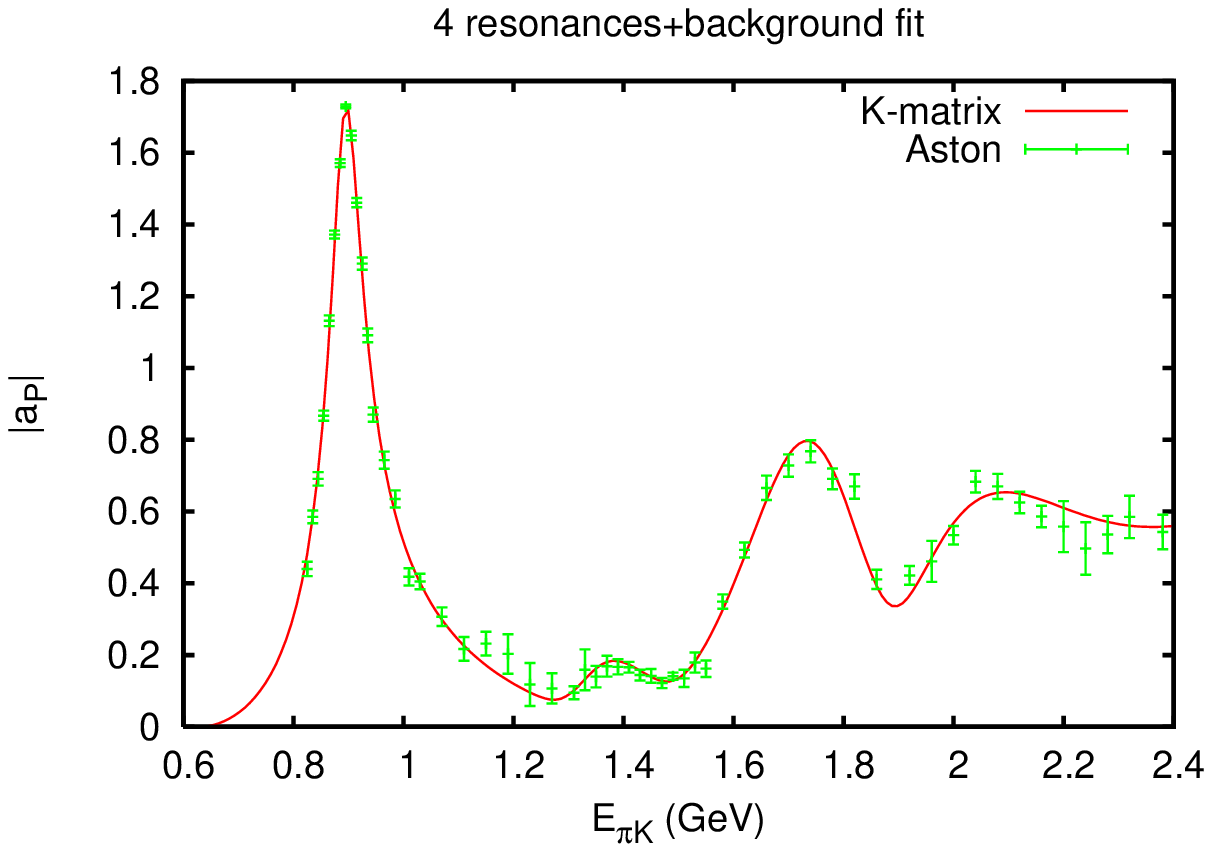}\\
\includegraphics[width=9cm]{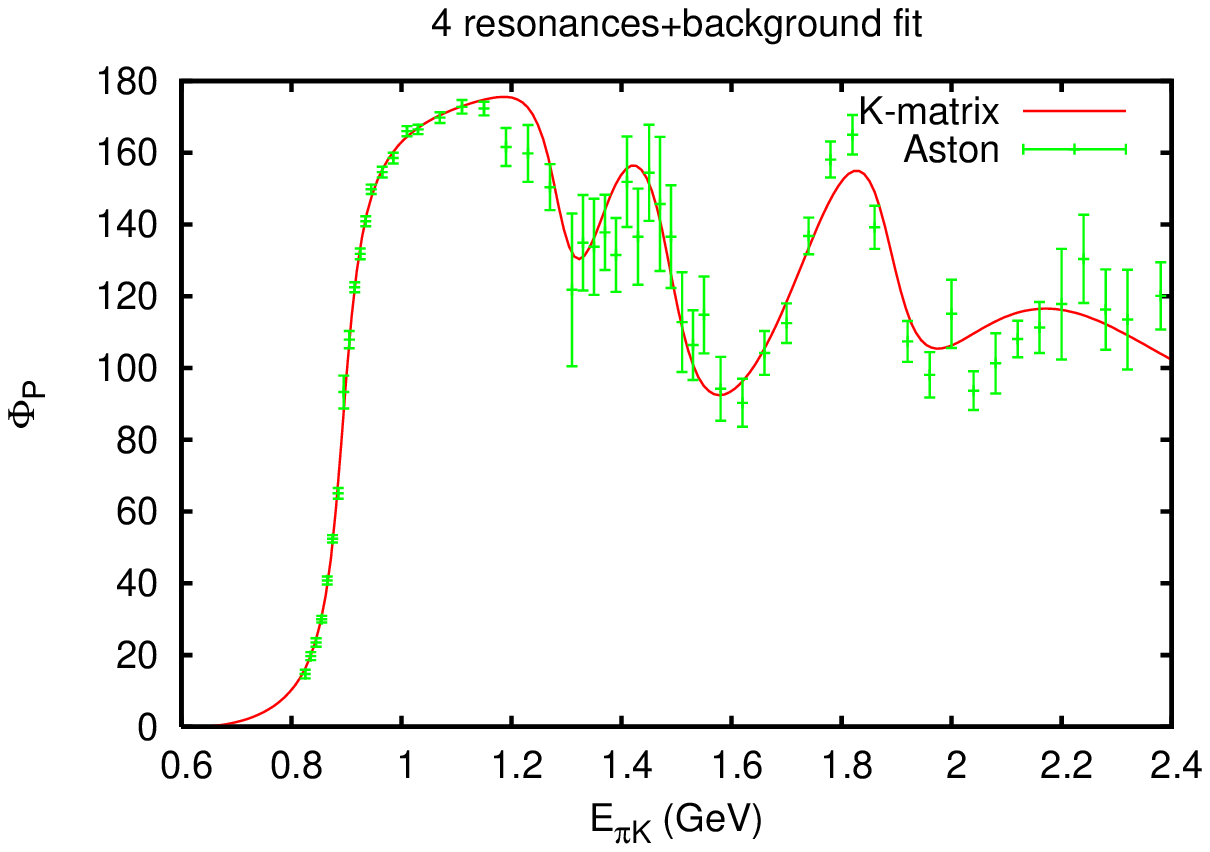}
\caption{\sl $K$-matrix fit to the $K \pi\to K \pi$ data of ref.\cite{aston88}
with four resonances plus background, as described in sec.~\sect{fitsophi}}
\lblfig{fitsophi}
\end{figure}

In the very small $s$ region finally, our $T$-matrix is expected to be 
qualitatively reasonable but certainly not very accurate because of the 
lack of constraints from experimental data in the threshold region. 
The $\pi K$ scattering length, for instance, is found to be
\be
a_1^{1/2}=0.025
\en 
which has the correct order of magnitude but is slightly larger
than the result from ChPT (at order $p^6$)\cite{bijdhonte}:
$a_1^{1/2}=0.018$.

\begin{figure}[ht]
\centering
\includegraphics[width=12cm]{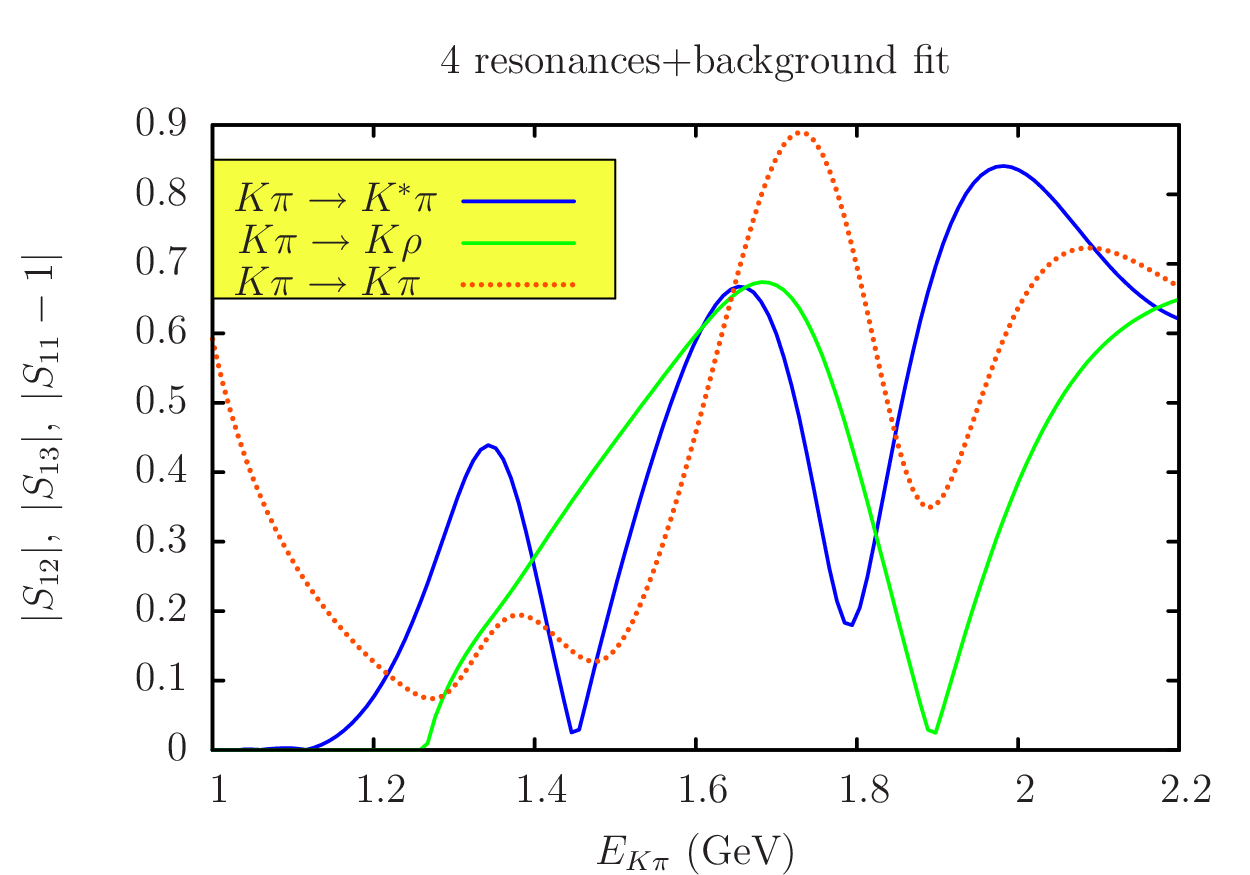}\\
\caption{\sl Results from the fit described in sec.~\sect{fitsophi}
concerning the inelastic $S$-matrix elements moduli $\vert S_{12}\vert$ 
and $\vert S_{13}\vert $ 
as well as the elastic matrix element modulus $\vert 1- S_{11}\vert$. }
\lblfig{Sij}
\end{figure}

\section{The $K\pi$ vector form factor from $n$-channel MO equations}
\subsection{Unitarity equations}
The Muskhelishvili-Omn\`es equations derive from the fact that, firstly,
each of the form factors $H_1(t)$, $H_2(t)$ and $H_3(t)$ satisfy
an unsubtracted dispersion relation because they are analytic 
functions in the variable $t$ except for a right-hand cut 
\footnote{In reality, the form factors $H_2$, $H_3$ could display
anomalous thresholds since they involve an unstable particle. We ignore this
possibility in our analysis.}~\cite{barton} and they decrease for $t\to\infty$
faster than $1/t$. Secondly, one can express the imaginary parts of the
form factors in terms of $T$-matrix elements using time-reversal 
invariance. Let us briefly repeat the derivation in the case of a
scalar operator ${\cal O}$ which is 
${\cal T}$-invariant
\be\lbl{tot}
{\cal T} {\cal O} {\cal T}^{-1}= {\cal O}\ .
\en
We consider the imaginary part of the 
matrix element between the vacuum and a state $\vert m>$ which we take 
to be an ``out'' state
\be\lbl{imagpart}
2i\, \im \outleft{m} \vert {\cal O}\vert 0> = 
\outleft{m} \vert {\cal O}\vert 0> -\outleft{m}\vert {\cal O}\vert 0>^*
\en
Using the fact that ${\cal T}$ is antiunitary and transforms an ``out''
sate into an ``in'' state, it is not difficult to 
transform the second term on the right-hand side in \rf{imagpart} and
recover the usual unitarity expression for the imaginary part
\be\lbl{unitequat}
\im \outleft{m}\vert {\cal O}\vert 0> = {1\over2} \sum_n T_{mn}^* 
\outleft{n}\vert {\cal O}\vert 0> \ .
\en 
In reality, we will use operators which carry space-time 
indices and thus are not ${\cal T}$-invariant. 
It is however easy to see that the
${\cal T}$ variation of the operator is compensated by 
the ${\cal T}$-variation of the
momenta $p_i$ in the state $\vert m>$ such that eq.~\rf{unitequat} remains
valid for the form factors.
For a given energy of the state $\vert m>_{out}$ 
a finite number of states $\vert n>_{out}$ 
contribute to the sum in the right-hand
side (Watson's theorem~\cite{watsonth} 
follows if the energy is lower than the first
inelastic threshold). If we truncate the summation in eq.~\rf{unitequat}
and insert the imaginary parts into the dispersion 
relations we obtain a closed set of MO integral equations for the form 
factors.  

\subsection{Application to $K\pi$, $K^*\pi$, $\rho K$ vector form factors}
We start from the general expression~\rf{unitequat}, apply it to the
operator $\bar{u}\gamma^3 s$ and retain 
three states in the sum: $K\pi$, $K^*\pi$, $\rho K$. We then use the 
formulas~\rf{j3kpi}~\rf{j3vp} and the 
partial-wave expansions~\rf{Tpwexp} of the $T$-matrix elements and 
compute the phase-space integrals. As expected, only the $J=1$  partial-wave
contributes and, after a small calculation the unitarity equations 
for $H_1(t)$, $H_2(t)$ and $H_3(t)$  are obtained. 
They can be written as follows, in matrix form
\be\lbl{Puniteq}
\im
\left(  \begin{array}{c}
H_1 \\
H_2 \\
H_3 \\
\end{array}\right)
= \tau^{-1} T^*\, Q^2\, \Sigma\, \tau 
\left(  \begin{array}{c}
H_1 \\
H_2 \\
H_3 \\
\end{array}\right)
\en
where $T$ is the 3x3 $J=1$ $T$-matrix, the diagonal matrices $Q^2$
and $\Sigma$ are defined in eq.~\rf{qetsigdef} and $\tau$ is also
a diagonal matrix
\be
\tau= diag\,(1,\sqrt{t},\sqrt{t})\ .
\en
The MO integral equation set then derives by combining eqs.~\rf{Puniteq}
with the unsubtracted dispersion relations satisfied by the form
factors
\be\lbl{disprel3}
\left(  \begin{array}{c}
H_1(t) \\
H_2(t) \\
H_3(t) \\
\end{array}\right)
=
{1\over\pi}\int_{(m_K+m_\pi)^2}^\infty 
{dt'\over t'-t} \im
\left(  \begin{array}{c}
H_1(t') \\
H_2(t') \\
H_3(t') \\
\end{array}\right)
\en
We recall here that the following matrix plays a role when discussing 
existence and multiplicity of the 
solutions to the MO equations\cite{muskhebook,moiL6}
\be
\tilde{S}=1 + 2i\tau^{-1} T  Q^2 \Sigma \tau 
\en 
This matrix differs from the \smatrix,
as defined in eq.~\rf{smatdef}, but the determinants of the 
two matrices are equal
\be
\det{\tilde{S}}=\det{S} \ .
\en
\subsection{Asymptotic conditions on the $T$-matrix}\lblsec{asytmat}
The MO equations obeyed by the form factors are coupled, homogeneous
singular integral equations with a kernel linear in the $T$-matrix. 
The mathematical properties of such equations are exposed in 
Muskhelishvili's book\cite{muskhebook}. In particular, the number
of independent solutions $N$ is given by the index of the integral
operator which can be expressed in terms of the sum of the eigenphases
$\delta_i(t)$ of the $S$-matrix. For an $n\times n$ $S$-matrix,
\be\lbl{index}
\sum_1^n\, [ \delta_j(\infty) - \delta_j(0) ] = N \pi\ .
\en
In order to determine the form factors from the integral equations
one  must therefore impose $N$ independent conditions. 
In our case one has $n=3$ and 
the \smatrix has been constrained from experimental input
up to $E_0\simeq 2.5$ GeV. 
We make a key assumption here that $E_0$ is sufficiently large that the
asymptotic regime for the $T$-matrix sets up for values of $E$ not much
larger than $E_0$. At $E=E_0$ the sum of the eigenphases 
is $\sum\,\delta_i(E_0)\simeq 3.5 \pi$. 
We therefore expect the index $N$ to be either 3 or 4. 
We choose to adopt
an asymptotic condition which has an index $N=4$. This will enable us
to impose four conditions on the form factor.
Three of these use the values at the origin of the
form factors, $H_i(0)$. We will discuss below how these values are 
constrained by experiment.
As a fourth condition, we can enforce the
behaviour of $f_+(t)$ at infinity. In QCD, ignoring flavour symmetry
breaking, one should have\cite{brodskylepage,duncanmueller} 
\be\lbl{asyqcd}
\left.f_+(-Q^2)\right\vert_{Q^2\to\infty} 
\sim{16\pi\sqrt2\, \alpha_s(Q^2) F_\pi^2
\over Q^2 }
\en
We do not attempt to reproduce the logarithmic running of $\alpha_s$, 
and actually use eq.\rf{asyqcd} with a constant value $\bar\alpha_s=0.2$.
The condition \rf{asyqcd} is then easily implemented in the form of a sum rule
\be\lbl{asysumr}
{1\over\pi}\int_{(m_K+m_\pi)^2}^\infty dt' \im H_1(t')= 16 \pi \sqrt{2} 
\bar\alpha_s\,F_\pi^2\ .
\en
In the region
$t \ge t_0$ one must use a parametrization which respects the unitarity
of the \smatrix. For this purpose, we write the \smatrix in
exponential form,
\be
S= \exp(2i H )
\en
where $H$ is a real, symmetric matrix. We then define the interpolation
on the matrix $H$ such that
\bea
&& \lim_{t\to\infty} H_{12}(t),\ H_{13}(t),\ H_{23}(t) =0,\quad
\nonumber \\
&& \lim_{t\to\infty} H_{11}(t)+ H_{22}(t)+H_{33}(t)=4\pi\ ,
\ena
using the simple interpolating function
\bea
&& H_{ij}(t)= \left({t_0\over t}\right)^\gamma \bigg(
\alpha_{ij} + \beta_{ij} \left({t_0\over t}\right)^\gamma \bigg)+
\nonumber\\
&& \phantom{H_{ij}(t)=\left({t_0\over t}\right)^\gamma}
 H_{ij}(\infty) \bigg(1-\left({t_0\over t}\right)^\gamma \bigg)^2\ .
\ena
The parameters $\alpha_{ij}$ and $\beta_{ij}$ are determined such as
to ensure continuity of $H_{ij}(t)$ and its first derivative at $t=t_0$.
Consistent with the assumption made above on the setting of the 
asymptotic regime, the parameter $\gamma$ must be larger than one:
in practice, we will take $\gamma=2$. 
The condition on the trace of $H$ still leaves some freedom as to the
behaviour of each diagonal element. 
We will consider a plausible scenario in which the asymptotic eigenphases
satisfy
\be\lbl{scenarioc}
\delta_1(\infty)=2\pi,\quad \delta_2(\infty)=\delta_3(\infty)=\pi
\en
where $\delta_1$  is the largest eigenphase at $t=t_0$. The three
eigenphases are shown in fig.~\fig{eigenphases}.
\begin{figure}[htb]
\centering
\includegraphics[width=11cm]{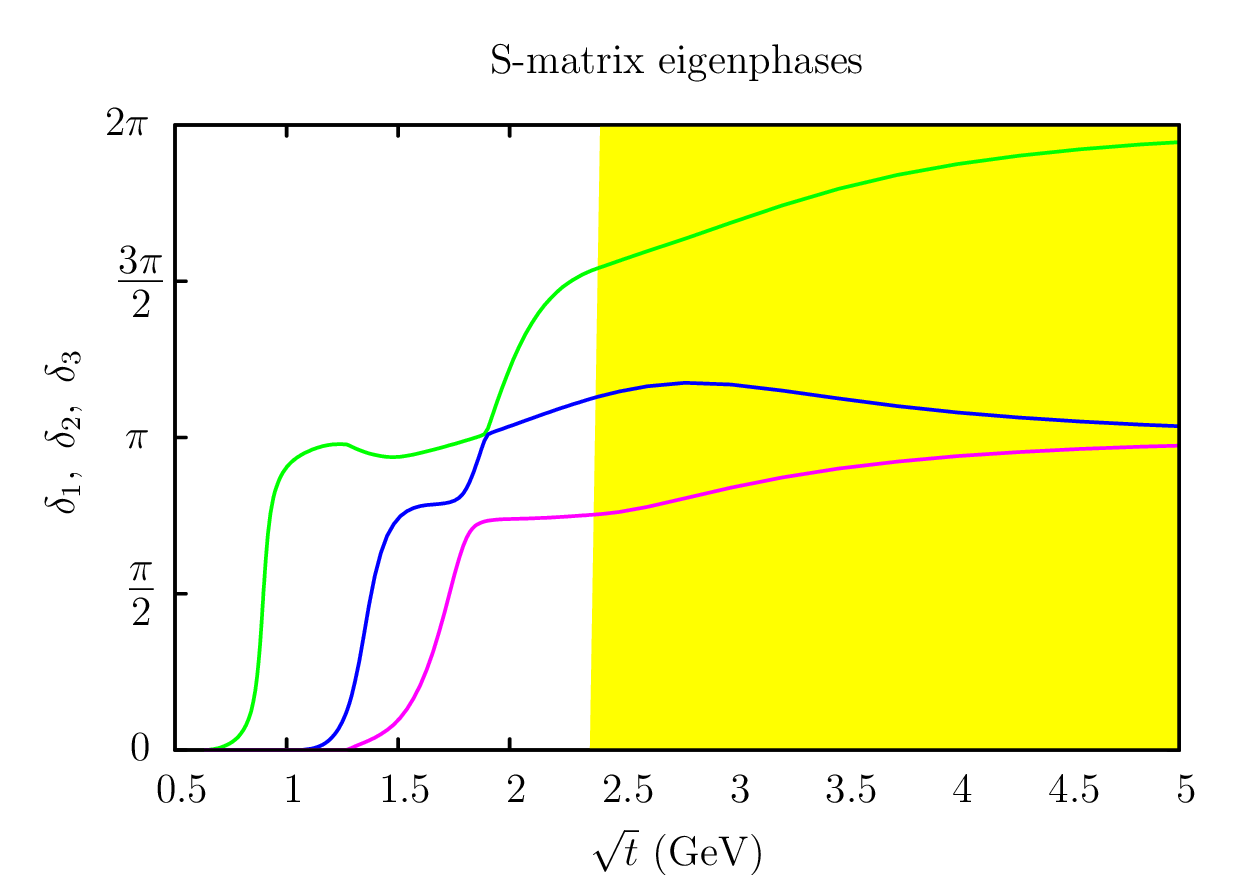}
\caption{\sl Eigenphases of the $S$-matrix: the left part of the 
figure shows the region determined from experiment, the right part (colored in
yellow) shows the asymptotic interpolation region. The asymptotic condition
here is $\sum\delta_i(\infty)=4\pi$.}
\lblfig{eigenphases}
\end{figure}

\subsection{Conditions at $t=0$}
In this section we discuss the experimental constraints on the
form factor components at $t=0$. 
The component $H_1(0)=f_+(0)$, firstly, is well known
from ChPT to very close to 1, the result at $O(p^4)$ was computed
in ref.~\cite{gl85ff},
\be\lbl{G10}
f_+(0)= 0.977 \quad  (ChPT\ O(p^4))\ .
\en 
We  note that ChPT computations at order $p^6$ have been performed\cite{bijkl3}
but the $p^6$ coupling constants involved are not yet well known.  
The remaining two form factor components involve vector mesons. We will argue
that their values at $t=0$ can still be evaluated reasonably well by appealing,
in addition to chiral expansion arguments, to a leading large $N_c$ 
approximation. In the chiral limit, flavour symmetry is exact, and the
following  relation holds between a charged current matrix element
and an electromagnetic current one,
\be
\braque{ K^{*\,+} \vert \bar{u}\,\gamma^\mu\, s\vert \pi^0}=
{3\sqrt{2}\over2} \braque{ \rho^+\vert j^\mu_{EM} \vert \pi^+}\ .
\en
This relation implies that one can relate $H_2(0)$ and $H_3(0)$
to the radiative width of the charged $\rho$ meson,
\be
\vert H_2(0)\vert=\vert H_3(0)\vert=
\left(  {27\over2}{\Gamma_{\rho^+\to\gamma\pi^+}\over\alpha\, p_{\gamma\pi}^3 }
\right)^{1\over2} \ .
\en
Using the experimental value~\cite{pdg06} of the radiative decay width
\be\lbl{exprop}
\Gamma_{\rho^+\to\gamma\pi^+} = 68\pm7\ {\rm keV}
\en
gives
\be\lbl{H2H30}
\vert H_2(0)\vert=\vert H_3(0)\vert = 1.54\pm0.08\ {\rm GeV^{-1}}\ .
\en
This method does not fix the sign of $H_2(0)$. For this purpose, we can
appeal to a simple vector-dominance picture applied, e.g., to 
the vertex function
$\braque{\gamma \vert\bar{u}\gamma_\mu d \vert\pi^-  }  $. In such a picture
$H_2(0)$ gets related to the ABJ anomaly\cite{abj},
\be\lbl{H2H30b}
H_2(0)=-H_3(0)= {N_c\over 8\pi^2}\, {1\over \sqrt2 f_V F_\pi}
\simeq 1.50\ {\rm GeV^{-1}}\ .
\en
We observe that the absolute value of $H_2(0)$ 
in the VMD model is in rather good
agreement with the one deduced from experiment.

We can try to refine these estimates by taking into account the breaking
of flavour symmetry to first order in the quark masses. For this purpose,
let us write down an effective Lagrangian,
\bea\lbl{lagsprad}
&& \Lag =\epsilon_{\mu\nu\alpha\beta}  \Big\{
   \ h_V\,\tr (V^\mu\{u^\nu,f_+^{\alpha\beta}\}) 
\nonumber \\
&& \phantom{ \Lag =\epsilon_{\mu\nu\alpha\beta}  }
+{a\over M^2_V}\, \tr([V^\mu,u^\nu][\chi^+,f_+^{\alpha\beta}]) 
\nonumber \\
&& \phantom{ \Lag =\epsilon_{\mu\nu\alpha\beta} }
+{b\over M^2_V}\, \tr(\{V^\mu,u^\nu\}\{\chi^+,f_+^{\alpha\beta}\})        
\nonumber \\
&& \phantom{ \Lag =\epsilon_{\mu\nu\alpha\beta}  }
+{c\over M^2_V}\, \tr([V^\mu,\chi^+][u^\nu,f_+^{\alpha\beta}])\Big\} \ .
\ena
We have used the same notation as in ref.~\cite{prades93} for the first term
in this Lagrangian. 
Flavour symmetry
breaking effects can be encoded into three independent Lagrangian terms:
this holds in the leading large $N_c$ approximation (since
multiple trace terms are suppressed). We can then constrain the coupling
constants $h_V$, $a$, $b$ and $c$ by considering the
radiative decays $\rho^+\to\pi^+\gamma$,  
$K^{*\,+}\to K^+\gamma$, and  $K^{*\,0}\to K^0\gamma$. The corresponding
amplitudes, computed from the Lagrangian~\rf{lagsprad}, have the
expressions
\be
A(\rho^+\to \pi^+\gamma)= {4\sqrt2\over3 F_\pi} h_V^{eff},\ 
\ h_V^{eff}= h_V +{4\mpid\over M_V^2}\,b 
\en
and
\bea
&& A(K^{*\,+}\to K^+\gamma)= {4\sqrt2\over3 F_K} h_V^{eff}
-{16\sqrt2\over 3F_\pi} {\mkd-\mpid\over M_V^2}(2b-3c)\nonumber\\
&& A(K^{*\,0}\to K^0\gamma)={-8\sqrt2\over3 F_K} h_V^{eff}
-{32\sqrt2\over 3F_\pi} {\mkd-\mpid\over M_V^2} b\ .
\ena
The following experimental results are available~\cite{pdg06} for
the $K^*$ mesons radiative decays,
\bea
&& \Gamma( K^{*\,+}\to K^+\gamma)= 116 \pm 12\ {\rm keV},
\nonumber \\
&& \Gamma( K^{*\,0}\to K^0\gamma)= 50\pm 5 \ {\rm keV} \ ,
\ena
which, together with the result~\rf{exprop} allows one to determine
three coupling-constants,
\bea\lbl{numvals}
&& h_V^{eff}= 0.0356\pm 0.0018,\nonumber \\ 
&& b=0.0010\pm 0.0015,\nonumber \\ 
&& c=0.0032\pm 0.0013\ .
\ena
Let us now compute our form factor components from the 
Lagrangian~\rf{lagsprad}, obtaining
\bea
&&  H_2(0)= {4\over F_\pi} h_V^{eff} -{16\over F_\pi} {\mkd-\mpid\over M_V^2}
(a-b+c)\nonumber\\
&&  H_3(0)= -{4\over F_K} h_V^{eff}  -{16\over F_\pi} {\mkd-\mpid\over M_V^2}
(a+b)\ .
\ena
These expressions show that $H_2(0)$ and $H_3(0)$ depend on the symmetry
breaking parameter $a$ which is left undetermined by the analysis of radiative
vector meson decays. Numerically, using the values \rf{numvals} we find
\bea
&& H_2(0)= 1.41 \pm 0.09 -65.4\, a\ {\rm GeV^{-1}},\nonumber\\
&& H_3(0)= -1.34 \pm 0.07 -65.4\, a\ {\rm GeV^{-1}}\ .
\ena 
We expect 
the parameter $a$ to have the same order of magnitude as the
other two symmetry breaking parameters, i.e. $\vert a\vert < 10^{-2}$.

\section{Solutions and comparisons with experimental results}
\subsection{Solving the MO equations}
The integral equations~\rf{Puniteq}~\rf{disprel3} can be solved numerically
by discretizing the integrals. This must be done in a way which guarantees
a precise evaluation of the principal-value integrals. For this purpose,
we use expansions over Legendre polynomials and exact expressions for the
principal-value integrals of these. More details can be found in 
ref.~\cite{moiL6}. One then arrives at a set of $M\times M$ homogeneous
linear ordinary equations. The Muskhelishvili conditions on the number of
solutions imply that the determinant of the system must vanish (if it
does not, then the only solution would be the trivial identically vanishing one)
and the associated matrix must have $N$ (with $N=4$ in our case) 
zero eigenvalues. In practice, because of discretization and roundoff errors 
no eigenvalue vanishes exactly, but one does have $N$ eigenvalues which are
very small in magnitude. A numerically stable way to proceed is to 
enlarge the system to an $(M+N)\times M$ one by adding $N$ constraints
on the form factors
as equations and then solve the new system using the singular value 
decomposition method. We obtained precise results with $M\gapprox 100$ 
and we used values of $M$ up to 400. Several correctness and accuracy tests
can be performed. In particular, while exact solutions are not known for
general $T$-matrices, the value of the determinant of the matrix formed
from $N$ independent solutions can be expressed in analytical
form\cite{muskhebook,moiL6} in terms of $S$-matrix elements.

\subsection{Results for $\tau$ decays}
The vector form factor $f_+(t)$ can be probed using the $\tau$ decay mode
$\tau\to K\pi \nu_\tau$. The energy distribution of the $K \pi$ pair 
has the following expression which involves $f_+(t)$ as well as the
scalar form factor $f_0(t)$
\bea\lbl{tauspeckpi}
&&{d\Gamma_{K\pi}(t)\over d\sqrt{t}} =
{V_{us}^2 G_F^2 m_\tau^3 \over 128 \pi^3} {q_{K\pi}(t) }
\left( 1- {t\over \mtaud} \right)^2 \times\\
&& \left[ 
\left( 1+ {2t\over \mtaud}\right) 
{4q^2_{K\pi}(t)\over t}     \vert f_+(t)\vert^2\right.
\left.
+{3(\mkd-\mpid)^2\over t^2} \vert f_0(t)\vert^2
\right]\ ,\nonumber
\ena
where the definition of $f_0$, in terms of the form factors 
introduced in eq.~\rf{def1} is
\be
f_0(t)=\sqrt2\left[  f_+^{K^+\pi^0}(t)  +{t\over\mkd-\mpid}
f_-^{K^+\pi^0}(t)\right]\ .
\en
The contribution of $f_0$ is kinematically suppressed except for very small
values of $t$. 
Let us begin by fixing the symmetry breaking parameter $a$
from the integrated width. Assuming exact iso\-spin symmetry one has
\be
\Gamma_{K\pi}= 3 \Gamma_{K^-\pi^0}={3\over2}\Gamma_{\overline{K}_0\pi^-}\ .
\en
The following value for the rate is quoted by the PDG\cite{pdg06}
\be\lbl{pdgrate}
R^{PDG}_{K\pi}= (13.5\pm0.05)\,10^{-3}\ .
\en
The most recent results from the Babar and Belle 
collaborations\cite{babarkpi,bellekpi}tend, however, to point towards 
a slightly smaller value 
\bea
&& R^{Babar}_{K\pi}=(12.48\pm0.009\pm0.054)\,10^{-3},\quad
\nonumber\\
&& R^{Belle}_{K\pi}=(12.12\pm0.006\pm0.039)\,10^{-3} 
\ena
(assuming isospin symmetry). These results can be reproduced 
with our vector form factor\footnote{In applications to $\tau$ decays
we use for $V_{us}$ the value such that $V_{us} f_+(0)=0.2167$  
quoted by the averaging group~\cite{moulson}. } 
by having the parameter $a$ in the range
\be
a= (-7.0^{+0.7 }_{-2.0} )\, 10^{-3}\ .
\en 
The central value corresponds to the PDG result and
the value $a=-9.0\,10^{-3}$ reproduces Belle's central 
figure for the rate. 
The energy distribution in the decay $\tau\to K\pi \nu_\tau$ has been measured
for the first time by the Aleph collaboration\cite{aleph99}.
The Belle collaboration has recently measured the distribution in energy 
for the decay $\tau^-\to K^0_S \pi^-\nu_\tau$  with 
considerably higher statistics~\cite{bellekpi} 
(approximately a thousand times larger). 
The Babar collaboration has also presented results for the energy
distribution for $\tau^-\to K^-\pi^0\nu_\tau$~\cite{babarkpi} 
but the data corrected for background are not publicly available. 
Data from Babar on $\tau^-\to K^0_S \pi^-\nu_\tau$ 
have been analyzed in a thesis~\cite{andrewlyon}
but the results have not yet been published.
\begin{figure}[ht]
\centering
\includegraphics[width=12cm]{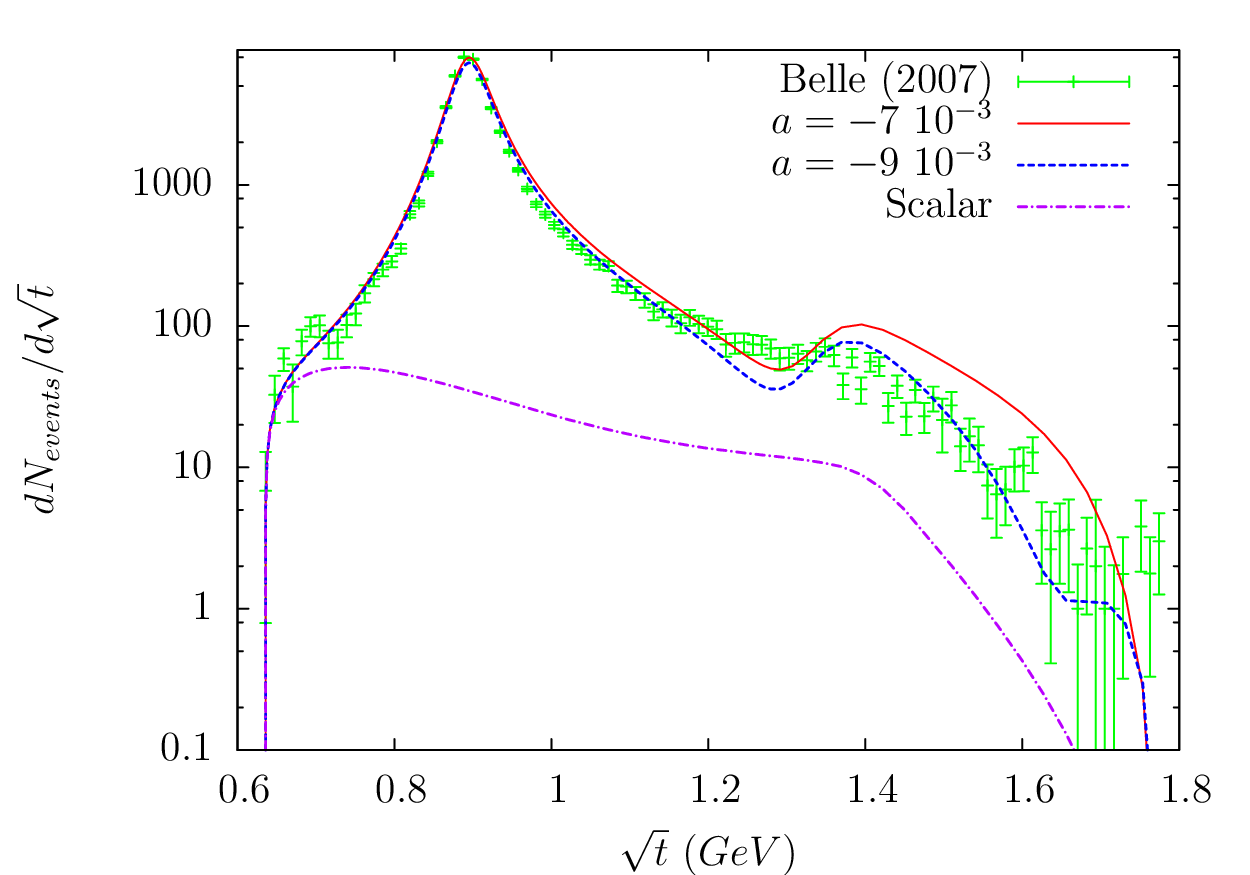}
\caption{\sl Results for the energy dependence of the $\tau$ decay rate
$\tau\to K \pi \nu_\tau$ using the vector form factor as discussed in the text
for two values of the symmetry breaking parameter $a$ compared with the
result from Belle~\cite{bellekpi}. The contribution from the scalar form
factor $f_0$ is shown as a dash-dot curve.
}
\lblfig{bellecomp}
\end{figure}

The result from our construction
is compared with the result of Belle\footnote{
The data shown have been corrected for 
background but not for acceptance, which we have assumed
to be approximately energy independent.}
in fig.~\fig{bellecomp}
for two values of $a$. In addition to the (dominant) contribution from the
vector form factor, we have also included the contribution from the scalar
form factor which we computed following ref.~\cite{jop}.
The parameter $a$ affects the size of the $K^*(892)$ peak (which controls
essentially the value of the integrated decay rate) and also the shape
of the form factor in the inelastic region. We find that it is possible to
reproduce both the integrated decay rate  and the shape of the 
energy distribution above the $K^*$ mass reasonably well. 
The solutions in the region $\sqrt{t}\gapprox 1.4$ GeV 
are sensitive to the assumptions
made on the $S$-matrix in the asymptotic domain. For instance, if we choose
an asymptotic condition with $N=3$ rather than $N=4$ we cannot get 
agreement with experiment when $\sqrt{t}\gapprox 1.4$ GeV by varying $a$.

It is instructive  to compare also our result for the vector form factor
with that obtained by Belle from a fit to their data. This comparison
avoids the problem of the acceptance (which is taken into account
in their fit) but one must keep in mind that there is some model dependence
in this determination (a model independent separation of the vector
and scalar form factors requires to analyze angular distributions). 
In ref.~\cite{bellekpi} five different fits, using descriptions 
of the form factors $f_+$, $f_0$ 
\`a la K\"uhn and Santamaria~\cite{kuhnsanta}, have been performed. 
In fig.~\fig{bellefit} we compare our result for the modulus of the
vector form factor with the experimental 
determination based on the fit which includes two vector
resonances: $K^*(892)$, $K^*(1410)$ for the vector form factor 
and one scalar resonance: $K^*_0(800)$ for the scalar form factor.
For clarity, we have presented our model's result as one curve corresponding
to the best fit to the LASS data but, obviously, one should keep in mind 
the uncertainties of these data, which are sometimes sizable, e.g., in
the 1.3 GeV region (see fig.~\fig{fitsophi} ).  

\begin{figure}
\centering
\includegraphics[width=12cm]{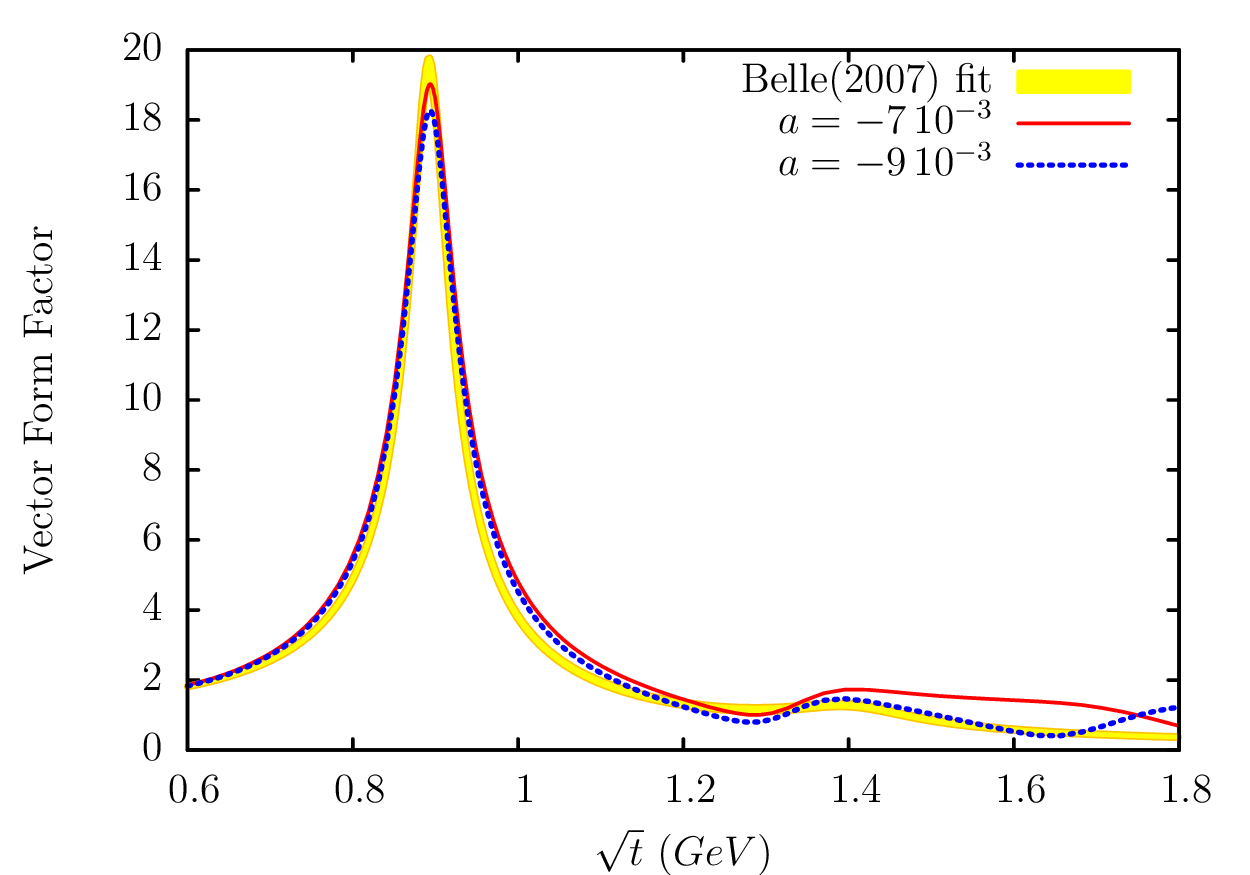}
\caption{\sl The yellow band represents the modulus of the vector form
factor determined from a four parameter fit in ref.~\cite{bellekpi}. It is
compared with our determination for two values of the symmetry breaking 
parameter $a$.
}
\lblfig{bellefit}
\end{figure}
This comparison deserves a comment. 
In the $K^*(892)$ region one expects to observe some isospin
symmetry breaking effects since the $K \pi$ system in $\tau\to K\pi\nu_\tau$ is
in a charged state while the experimental data from LASS concern $K \pi$ 
in a neutral state. Fig.~\fig{bellefit} shows a visible difference concerning
the width which is narrower for the $K^{*+}$ than for the $K^{*0}$. Somewhat
surprisingly, no difference is seen concerning the mass. In fact, the 
mass resulting from Belle's fit: $M_{K^{*+}}= 895.47\pm 0.20\pm 0.44 
\pm 0.59$ differs by about 4 MeV from the mass quoted in the PDG:
$M_{K^{*+}}= 891.66\pm0.26$ which is based on hadronic production experiments.
This shift in the $K^{*+}$ mass was reported earlier by CLEO~\cite{cleo}
but remains to be confirmed. It could be a similar effect 
to the one observed for the $\rho$ meson mass.

The model also generates predictions for
$\tau$ decays into $K^*\pi$ and $K\rho$ via vector current. 
The energy distribution of the
decay width for $K^*\pi$ reads, in terms of the form factor $H_2$ 
\bea\lbl{gammvp}
&& {d\Gamma_{K^*\pi}\over d\sqrt{s}}=
\\
&& \quad {V_{us}^2 G_F^2 m_\tau^3 \over 32 \pi^3}    (q_{K^*\pi})^3
\left( 1- {t\over \mtaud} \right)^2 \left( 1+ {2t\over \mtaud}\right) 
\vert H_2(t)\vert^2\nonumber
\ena
and an exactly similar expression holds for $K\rho$ in terms of $H_3$. 
The results from
our construction are plotted in fig.~\fig{tauspectralvp}. The figure
shows that the $K^*(1410)$ appears very clearly in the $K^*\pi$ channel. 
The $\tau$ decay into $K\rho$ (via vector current) is comparatively
strongly suppressed. These features reflect the constraints which we
have used in the construction of the $T$-matrix based on the LASS
experiments. The results for the integrated rates are 
\bea
&&R(\tau\to K^* \pi \nu_\tau)_V= (1.37  \pm0.02   )\,10^{-3}
\nonumber\\
&&R(\tau\to K  \rho \nu_\tau)_V= (4.5   \pm3.0    )\,10^{-5}
\ena
where the errors are estimated by simply varying the parameter $a$. 
Our result for
$R(\tau\to K^* \pi \nu_\tau)_V$ is seen to agree 
with the one quoted by the Aleph collaboration~\cite{aleph99},
\bea
&&R^{Aleph}(\tau\to K^*(1410)\nu_\tau \to K\pi \pi \nu_\tau)
=\nonumber\\
&&\phantom{R^{Aleph}(\tau\to K^*(1410)}
(1.4^{+1.3\ +0.0}_{-0.9\ -0.4})\times 10^{-3}\ .
\ena
which, however, is not very precise.

\begin{figure}[ht]
\centering
\includegraphics[width=12cm]{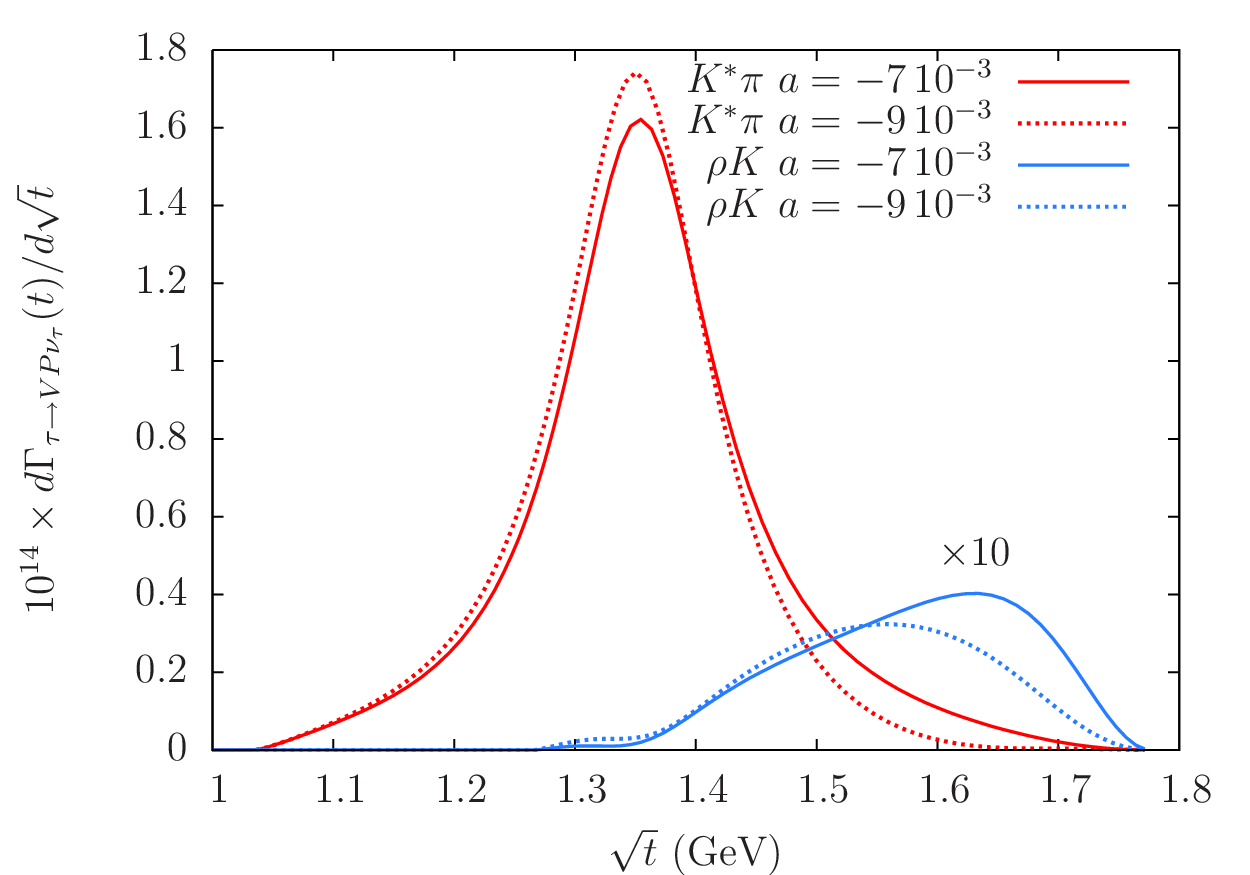}
\caption{\sl Results for the energy distribution of the $\tau$ decay width
into $K^*\pi$ and $K\rho$ (rescaled by a factor of 10) 
via vector current from our model, for two values 
of the symmetry breaking parameter $a$.
}
\lblfig{tauspectralvp}
\end{figure}

\subsection{Some results at $t=0$}
For values of $t$ near $t=0$, experimental results on $f_+(t)$ can be
obtained from $K_{l3}$ decays and several new experiments have been performed
recently. One usually defines the slope and curvature parameters from 
the Taylor expansion
\be\lbl{f+exp}
f_+(t)= f_+(0)\left( 1 + \lambda' {t\over m_{\pi^+}^2 } +{1\over 2}\,
\lambda'' {t^2\over m_{\pi^+}^4 } + ... \right)\ .
\en
The results which we get for $\lambda'$ and $\lambda''$ are 
\be\lbl{lambres}
\lambda'= \left(26.05^{+0.21}_{-0.58}  \right)\,10^{-3}          ,\quad 
\lambda''= \left(1.29^{+0.01}_{-0.04}  \right)\,10^{-3}
\en
where the errors, again, simply reflect the range of variation of the
parameter $a$ (the lower values correspond to $a=-9\,10^{-3}$). 
They compare reasonably well with the average (performed in 
ref.~\cite{moulson}) over the recent experiments
\be
\lambda'_{exp}=(24.8\pm 1.1  ) \,10^{-3}  ,\quad
\lambda''_{exp}=(1.61\pm 0.45  ) \,10^{-3}\ .
\en
The results \rf{lambres} can also be compared with the 
predictions recently presented in ref.~\cite{joptau}\footnote{We converted  
the numerical values taking into account 
that the expansion formula \rf{f+exp} is used in ref.~\cite{joptau} 
with $m_\pi=138$ MeV rather than $m_\pi=139.57$ MeV.},
\be
\lambda'=26.2\,10^{-3},\quad \lambda''= 1.37\,10^{-3}\qquad
(\hbox{ ref.~\cite{joptau}})\ .
\en 
\subsection{Some remarks on the asymptotic region}
When $t\to\infty$, the MO equations driven by the $T$-matrix behaving as 
discussed in sec.~\sect{asytmat} can be shown to imply that $t f_+(t)$
goes to a constant. We have constrained this constant, via eq.~\rf{asysumr},
in order to correctly reproduce QCD in an average sense. Fig.~\fig{ffactorasy}
displays the real part of the product $t f_+(t)$ obtained from the 
numerical solution of the equations. The figure also shows the asymptotic
QCD expectation (analytically continued to the timelike region). 
One sees that  $t f_+(t)$ indeed goes to a constant, but the
asymptotic behaviour sets in at fairly large values of $t$. A similar
feature was observed in the case of the pion vector form factor
which was discussed in ref.~\cite{donoghff}.
\begin{figure}[ht]
\centering
\includegraphics[width=12cm]{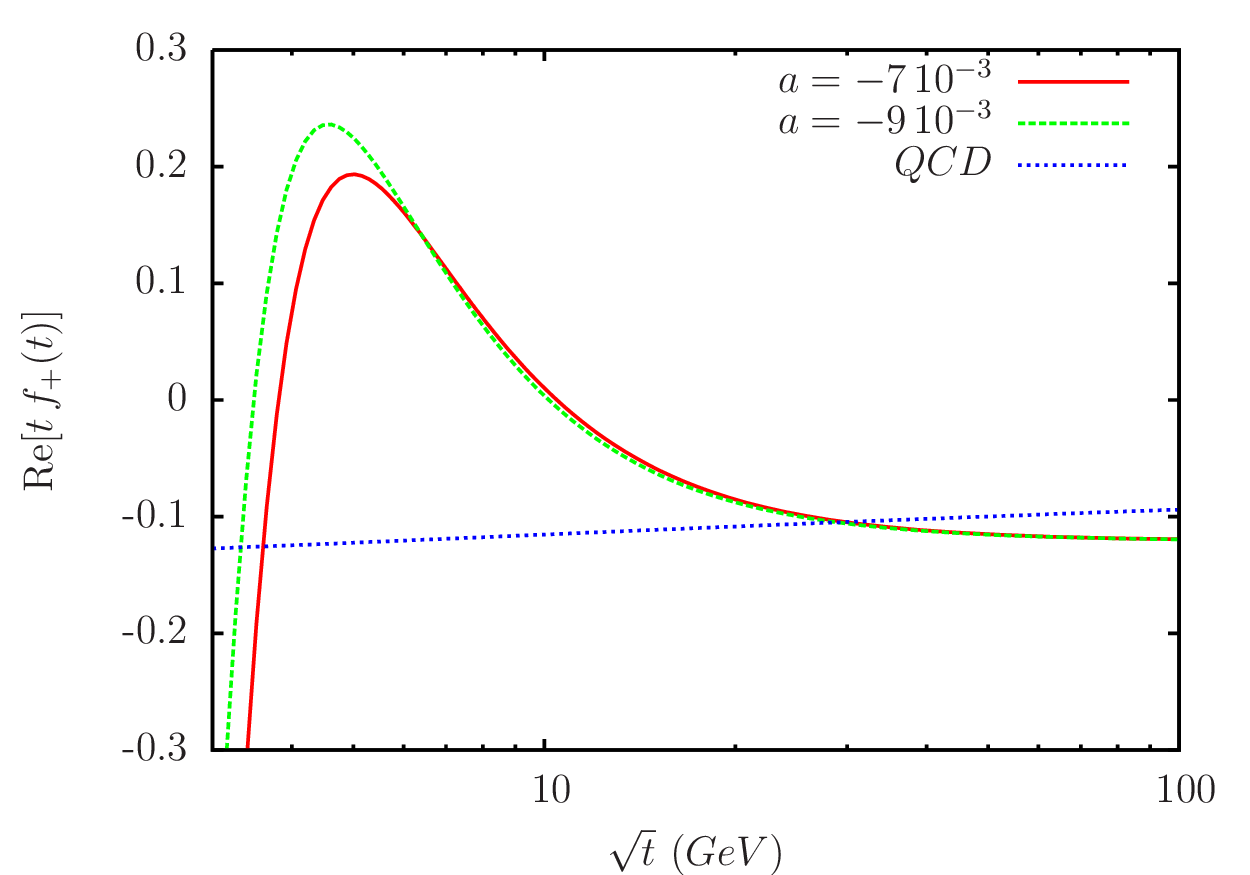}
\caption{\sl Real part of the product $t f_+(t)$ resulting from our
solution of the MO equations compared with asymptotic QCD prediction.
}
\lblfig{ffactorasy}
\end{figure}

Another remark concerns the phase of $f_+(t)$ 
(let us denote it as $\phi_+(t)$). It is often assumed that $\phi_+(t)$
should go to $\pi$ as $t$ goes to infinity (e.g.~\cite{yndurain}), 
while compatibility with asymptotic QCD simply implies 
that it should go to $\pi$ modulo $2\pi$. 
Let us recall that if the phase
of $f_+(t)$ goes to $\pi$ at infinity, $f_+$ can be expressed in terms of
its phase as a minimal Omn\`es representation. In our construction, we
find that the behaviour of the phase at infinity depends on the value of $a$.
This is illustrated in fig.~\fig{ffphase}. For the central value of $a$
the phase actually goes to $3\pi$ at infinity. When $a$ gets slightly smaller, 
a transition occurs and the phase goes to $\pi$. 
Again here, the asymptotic behaviour is reached for rather large values 
of $t$. 
%

%
%
 
\begin{figure}[ht]
\centering
\includegraphics[width=12cm]{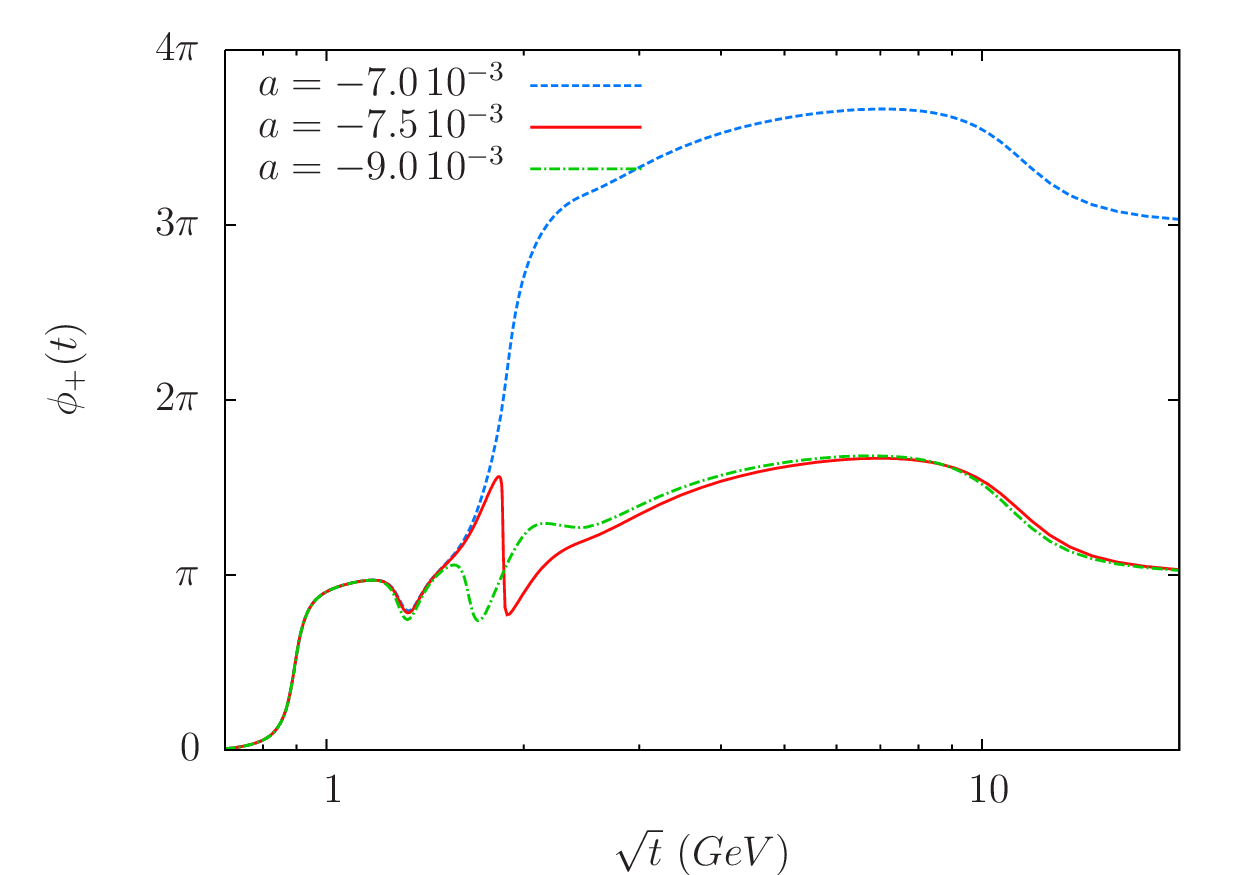}
\caption{\sl Illustration of the changing behaviour of the phase of $f_+(t)$ 
at large $t$ when the parameter $a$ varies.  
}
\lblfig{ffphase}
\end{figure}

\section{Conclusions}
In this work, we have considered matrix elements between light states of the 
stangeness changing vector current. Such quantities are now becoming 
accessible to experiment, with rather good precision in a rather large energy
range from hadronic $\tau$ decays. There are possible applications to
certain three-body B decays as well. 
From a theoretical point of view, one can establish relations between 
two-body form factors and the scattering matrix not only at very low energy
in the elastic scattering region (Watson's theorem) but also at somewhat
higher energies provided that inelastic scattering is dominated by 
two-body channels. This was shown to be the case for $\pi K$ scattering
in the $P$-wave by the LASS collaboration\cite{aston84,aston87,aston88} in
the energy range $E\lapprox 2.5$ GeV. Similar properties hold for
$\pi\pi$ scattering in the $S$-wave and also for 
$\pi K$ scattering in the $S$-wave and were used to construct scalar form
factors\cite{dgl,jop}. Those results, however, have not been compared
with experimental data. 

We have constructed a three channel $T$-matrix, which satisfies unitarity,
from fits to the large amount of data on $\pi K\to\pi K$ scattering
in the $P$-wave and satisfying the experimental constraints 
on inelastic scattering. In the region $E \ge 2.5$ GeV, 
one must make a plausible guess.  We impose a smooth
interpolation such that the MO operator has index $N=4$. 
The three form factors can then be determined by solving the system of MO
equations and applying four constraints. We have used the three values at
the origin, which we have argued to be determined, making use of chiral
and flavour symmetries, up to one symmetry breaking parameter $a$.  
As a fourth constraint, we used asymptotic QCD behaviour for the 
form factor $f_+$. 

The result of this construction for the form factor $f_+(t)$
compares reasonably well with the recent determination by the Belle 
collaboration. The agreement in the inelastic region is not quite trivial
to achieve. Indeed, the $K^*(1680)$ resonance appears as a very large effect
in $\pi K\to \pi K$   scattering while it is suppressed in the form factor. 
Varying the parameter $a$ outside of the range allowed from the
integrated $\tau\to K\pi\nu_\tau$ rate also destroys the agreement 
in the energy distribution. 
The result in the domain $\sqrt{t}\ge 1.4$ GeV is also sensitive to how the
$T$-matrix behaves above 2.5 GeV.
In the region of the $K^*(892)$ resonance, there are some discrepancies. 
While these could be due to isospin breaking effects, they seem to concern
the width rather than the mass of the $K^*(892)$, which is unexpected.
Finally, we made predictions for the total rates and the energy distributions
for the vector current contribution to the $\tau$ decays 
$\tau\to K^*\pi\nu_\tau$ and $\tau\to \rho K \nu_\tau$. 
\medskip

\noindent{\Large\bf Acknowledgements}\\[5pt]{
The author would like to thank D. Epifanov for making the numerical values
of the Belle measurements available and for correspondence. This work is
supported by the European commission MRTN FLAVIAnet [MRTN-CT-2006035482].
}

\end{document}